\documentclass[conference]{IEEEtran}
\IEEEoverridecommandlockouts
\usepackage{cite}
\usepackage{amsmath,amssymb,amsfonts}
\usepackage{algorithmic}
\usepackage{graphicx}
\usepackage{textcomp}
\usepackage{xcolor}
\def\BibTeX{{\rm B\kern-.05em{\sc i\kern-.025em b}\kern-.08em
    T\kern-.1667em\lower.7ex\hbox{E}\kern-.125emX}}
\begin{document}

\title{On Parametric Amplification In Discrete Josephson Transmission Line}

\author{\IEEEauthorblockN{Eugene Kogan}
\IEEEauthorblockA{\textit{Department of Physics
Bar-Ilan University
Ramat-Gan, Israel }\\
\textit{Donostia International Physics Center (DIPC)
San Sebastian/Donostia, Spain}\\
Eugene.Kogan@biu.ac.il}
}

\maketitle

\begin{abstract}
We consider the discrete series-connected lossy Josephson transmission line, constructed  from  Josephson junctions,  capacitors and resistors
(one-dimensional array of Josephson junctions).  We derive equations describing pump, signal, and idler interaction in the system and calculate the thresholds for the  parametric amplification.
\end{abstract}


\section{Introduction}
Superconducting parametric amplifiers attract a lot of interest, due to their importance
in microwave electronics  \cite{beltran,nation,eom,aumentado}.
Traditional amplifiers  comprise a
single Josephson junction (JJ) or an array of junctions in a resonant
cavity which ultimately limits the bandwidth and dynamic
range.

Recently, owing to impact of the
kinetic-inductance traveling-wave parametric amplifier, the Josephson traveling-wave parametric amplifiers
 enabling larger gain per unit length with
lesser pump power have been in the particular focus of
several research groups
\cite{yaakobi,brien,macklin,kochetov,bell,white,zorin,
fasolo,basko,dixon,zorin2,miano,reep,pekker,greco,yuan,kow,katayama}.

We studied previously \cite{kogan3} the problem of parametric amplification in the Josephson transmission line (JTL) in the continuum approximation.  Now we consider the problem for the discrete JTL.

\section{RSJ Josephson transmission line}

\begin{figure}[h]
\includegraphics[width=\columnwidth]{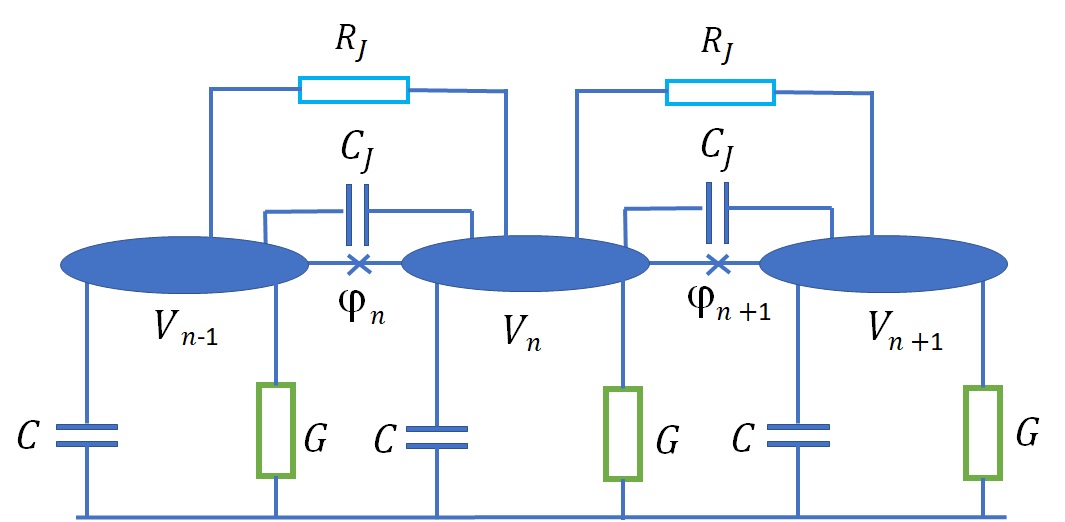}
\caption{Josephson transmission line  composed of superconducting grains}
 \label{gr1}
\end{figure}

We consider a  model of the JTL  presented in Fig. \ref{gr1}.

We take
as the dynamical variables  Josephson phases
 and the potentials of the grains $V_n$.  The circuit equations are  \cite{kogan3}
\begin{subequations}
\begin{alignat}{4}
&\frac{\hbar}{2e}\frac{d \varphi_n}{d t} =V_{n-1}-V_n ,\\
&C\frac{dV_n}{dt}+GV_n=\partial^2_n \left(\frac{V_{n}}{R_J}+C_J\frac{dV_n}{dt}\right)\nonumber\\
&+I_c\left[\sin\left(\varphi_n+\varphi_0\right)
- \sin\left(\varphi_{n+1}+\varphi_0\right)\right].
\end{alignat}
\end{subequations}
where    $C$ is the ground capacitance,  $I_c$ is the critical current of the JJ,
$G$ is the conductance of the ohmic resistor  shunting the ground
capacitor,  $R_J$ is the ohmic resistor shunting the JJ;
the second-order discrete
differential operator $\partial^2_n$ is defined as
$\partial^2_n f_n\equiv f_{n-1 }-2f_n+f_{n+1 }$.
For the sake of generality we consider d.c. background Josephson current $I_c\sin\left(\varphi_0\right)$ flowing along the JTL,
where $\varphi_0$ is the d.c. Josephson phase.

Inclusion of the capacitance in parallel to the Josephson elements (inter-island capacitance) would certainly make the model more physically realistic.  However it complicates the mathematics (there appears the fourth order derivative  in the final equation). In spite of this the parallel capacitance was included in our  previous paper dealing with parametric amplification in the JTL in the continuum approximation \cite{kogan3}. It turned out that such generalization of the model doesn't lead to any qualitative differences. So in the present paper we decided to give the priority to simplicity over generality.

We can exclude $V_n$  and obtain closed equation for $\varphi_n$
\begin{eqnarray}
\label{co}
\frac{d^2\varphi_n}{d\tau^2}+Z_JG\frac{d\varphi_n}{d\tau}\nonumber\\
=\partial^2_n \left[\sin\left(\varphi_n+\varphi_0\right)
+\frac{Z_J}{R_J}\frac{d\varphi_n}{d\tau}
+\frac{C_J}{C}\frac{d^2\varphi_n}{d\tau^2}\right],
\end{eqnarray}
where we  introduced the dimensionless  time $\tau=t/\sqrt{L_J C}$,
 $Z_J\equiv\sqrt{L_J/C}$ is the characteristic impedance of the JTL, and $L_J=\hbar/(2eI_c)$.
Let us present  (\ref{co}) as
\begin{eqnarray}
\label{cob}
\frac{d^2\varphi_n}{d\tau^2}-\partial_n^2\left(\cos\varphi_0\cdot\varphi_n
+\frac{C_J}{C}\frac{d^2\varphi_n}{d\tau^2}\right)\nonumber\\
=\partial_n^2\left\{\frac{Z_J}{R_J}
\frac{d\varphi_n}{d \tau}+n.l.[\sin\left(\varphi_n+\varphi_0\right)]\right\}
-Z_JG\frac{d\varphi_n}{d\tau},
\end{eqnarray}
where n.l.  stands for the non-linear terms of expansion of the sine function in Taylor series.

If we ignore the dissipation and the non-linear terms, Eq. (\ref{cob}) has obvious solution
\begin{eqnarray}
\label{combi}
\varphi_n=\frac{1}{2}\sum_{\alpha}A_{\alpha}e^{i(k_{\alpha}n-\omega_{\alpha} \tau)}+c.c.,
\end{eqnarray}
where
\begin{eqnarray}
\label{5}
\omega_{\alpha}^2=\frac{4\cos\varphi_0\sin^2\left(\frac{k_{\alpha}}{2}\right)}
{1+\frac{4C_J}{C}\sin^2\left(\frac{k_{\alpha}}{2}\right)},
\end{eqnarray}
and
 $A_{\alpha}$ are arbitrary constant amplitudes.
The  nonlinearity and the dissipation  we'll take into account approximately, by changing (\ref{combi}) to
\begin{eqnarray}
\label{combin}
\varphi_n=\frac{1}{2}\sum_{\alpha}A_{\alpha}(n)e^{i(k_{\alpha}n-\omega_{\alpha} \tau)}+c.c.
\end{eqnarray}
and assuming that
the complex amplitudes $A_{\alpha}(n)$  slowly change with $n$.
Additionally, because we  consider the dissipation and nonlinear terms
as being in some sense small,  while calculating the discrete second order derivative of the terms in the parenthesis in (\ref{cob})
 we will ignore the corrections which come from the $n$-dependence of the amplitudes, thus the discrete second derivative operator acting on a partial wave with the wave vector $k$ just multiplies this partial wave by  $-4\sin^2\left(\frac{k}{2}\right)$. For example,
\begin{eqnarray}
\label{harn2}
\frac{d}{d \tau}\left(\varphi_{n-1}-2\varphi_n-\varphi_{n+1}\right)\nonumber\\
=-2i\sum_{\alpha}\sin^2\left(\frac{k_{\alpha}}{2}\right)\omega_{\alpha}A_{\alpha}
e^{i(k_{\alpha}n-\omega_{\alpha} \tau)}+c.c.,
\end{eqnarray}

However, we'll do better while calculating  the discrete second derivative standing in the l.h.s. of (\ref{cob}).
We promote $n$,  as argument of the amplitudes, to the continuous variable $Z$  and approximate
\begin{eqnarray}
\label{aa}
A_{\alpha}(n\pm 1)=A_{\alpha}(Z)\pm\frac{\partial A_{\alpha}(Z)}{\partial Z}.
\end{eqnarray}
As the result, for the discrete second derivative we obtain
\begin{eqnarray}
\label{nlin}
\varphi_{n-1}-2\varphi_{n}+\varphi_{n+1}\nonumber\\
=-2\sum_{\alpha}
\sin^2\left(\frac{k_{\alpha}}{2}\right)A_{\alpha}e^{i(\omega_{\alpha} \tau-k_{\alpha}n)}+c.c.,\nonumber\\
+i\sum_{\alpha}\sin (k_{\alpha})
\frac{\partial A_{\alpha}}{\partial Z}e^{i(k_{\alpha}n-\omega_{\alpha} \tau)}+c.c.
\end{eqnarray}

\section{3 waves mixing}

\subsection{Coupled equations for the amplitudes}
\label{adm}

Now let us take into account the nonlinear terms.
Consider the case $\varphi_0\sim 1$  and $\left|\varphi_n\right|\ll 1$
Expanding the sine function in the r.h.s. of (\ref{co}) in Taylor series  and keeping the first two terms of the expansion we have
\begin{eqnarray}
\label{co3}
n.l.\left[\sin\left(\varphi_{n}+\varphi_0\right)\right]
=-\frac{\sin\varphi_0}{2}\varphi_{n}^2.
\end{eqnarray}
After we substitute (\ref{combi}) into (\ref{co3}), the general result for the quadratic term would be too complicated,  so
we'll consider a superposition of only three waves (pump,
signal, and idler), with the frequencies $\omega_p$, $\omega_s$ and $\omega_i$ respectively,  satisfying equation
\begin{eqnarray}
\omega_s+\omega_i=\omega_p.
\end{eqnarray}
Keeping only the terms which have
$\omega_{p,s,i}t$ time dependence  we obtain
\begin{eqnarray}
\varphi_{n}^2
=\frac{1}{2}A_sA_ie^{i[(k_s+k_i)n-\omega_pt]}
+\frac{1}{2}A_pA_i^*e^{i[(k_p-k_i)n-\omega_st]}\nonumber\\
+\frac{1}{2}A_pA_s^*e^{i[(k_p-k_s)n-\omega_it]}+c.c.
\end{eqnarray}

Finally, collective all the perturbative terms in (\ref{co}) we
obtain coupled equations for the  amplitudes
 \begin{subequations}
\label{oco1}
\begin{alignat}{4}
\frac{dA_p}{dZ}+\nu_pA_p&=i\mu_pA_sA_ie^{i\Delta k\cdot Z}  \label{pump}\\
\frac{dA_{s,i}}{dZ}+\nu_sA_{s,i}&=i\mu_{s,i}A_pA_{i,s}^*e^{-i\Delta k\cdot Z},  \label{pumpb}
\end{alignat}
\end{subequations}
where
\begin{subequations}
\label{hren}
\begin{alignat}{4}
\nu_{\alpha}&=\frac{\tan (k_{\alpha}/2)}{\omega_{\alpha}}
\left[Z_JG+4\sin^2\left(\frac{k_{\alpha}}{2}\right)\frac{Z_J}{R_J}\right]
\label{nu}\\
\mu_p&=\frac{2\sin\varphi_0\cdot \tan (k_p/2)}{\omega_p^2}
\sin^2\left(\frac{k_s+ k_i}{2}\right)\\
\mu_{s,i}&=\frac{2\sin\varphi_0\cdot \tan (k_{s,i}/2)}{\omega_{s,i}^2}
\sin^2\left(\frac{k_p-k_{i,s}}{2}\right)\\
\Delta k&=k_s+k_i-k_p .
\end{alignat}
\end{subequations}

Comparing Eqs. (\ref{oco1})  with the appropriate equations from our previous publication \cite{kogan3}, we see that
the basic equations systems  in the discrete
consideration and in continuum approximation are the same.
Only the parameters of the systems are different
(compare  (\ref{hren}) with the appropriate equations from our previous publication \cite{kogan3}).

\subsection{Weak  signal: threshold for the parametric amplification}
\label{para}

Let us solve Eq. (\ref{oco1}) in the small (relative to the pump) signal and idler approximation  ($ |A_s|,|A_i|\ll |A_p|$).  In this approximation the
equation for $A_p$ becomes decoupled from the other two and takes the form
\begin{eqnarray}
\label{pu0}
\frac{dA_p}{dZ}+\nu_pA_p=0,
\end{eqnarray}
with the obvious solution
\begin{eqnarray}
\label{pu}
A_p(Z)=e^{-\nu_pZ}A_p^0.
\end{eqnarray}

Treating the other two equations, we'll introduce
 the local approximation, by treating $A_p$ in the r.h.s. of Eqs. (\ref{pumpb}) as a constant.  In this approximation  the solutions  for $A_s,A_i$  are
\begin{eqnarray}
\label{soli}
\left(\begin{array}{c} A_s\\A_i\end{array} \right)=
 e^{-i\Delta k /2\cdot Z}\left(\begin{array}{c}\dots e^{KZ} \\\dots e^{K^*Z}\end{array} \right),
\end{eqnarray}
where dots stand for some constant amplitudes, and  $K$  is one of the roots of the characteristic polynomial
\begin{eqnarray}
\label{sol2}
P(K)\equiv K^2+(\nu_s+\nu_i)K\nonumber\\
+(\nu_s-i\Delta k/2)(\nu_i+i\Delta k/2)-\mu_s\mu_i\left|A_p\right|^2.
\end{eqnarray}
Parametric amplification of the signal takes place when
one of the roots have positive real part. The boundary between the parametric amplification and no parametric amplification we can find by demanding that $K$ for that boundary is purely imaginary. So,
after simple algebra, we obtain
the condition for the parametric amplification:
\begin{eqnarray}
\label{3}
\mu_s\mu_i\left|A_p\right|^2>\nu_s\nu_p\left[1
+\frac{\left(\Delta k\right)^2}{(\nu_s+\nu_i)^2}\right].
\end{eqnarray}
Looking at Eq. (\ref{3}) we realise that the momenta mismatch $\Delta k$ acts in some sense similar to the losses in the system. Both factors together define the threshold for the parametric amplification.

Though it's not completely consistent, we bring  back the coordinate dependence of $A_p$, given by Eq. (\ref{pu}),
and consider (\ref{3}) as a local condition for the parametric amplification.

\subsection{Small wave vectors  limiting case}

Consider  the limiting case $k_p,k_s,k_i\ll 1$. In this case $\Delta k=0$.  We assume additionally that $R_J=\infty$, hence $\nu_{\alpha}=\nu$ (doesn't depend upon $\alpha$). Hence Eqs. (\ref{pump}),  (\ref{pumpb})  take the form
 \begin{subequations}
\begin{alignat}{4}
\frac{dA_p}{dZ}+\nu A_p&=i\mu_pA_sA_i  \label{umpa}\\
\frac{dA_{s,i}}{dZ}+\nu A_{s,i}&=i\mu_{s,i}A_pA_{i,s}^*, \label{umpb}
\end{alignat}
\end{subequations}
where
 \begin{eqnarray}
 \label{mu}
 \mu_p=\mu_s+\mu_i.
 \end{eqnarray}
We'll  present the amplitudes as
\begin{eqnarray}
A_p=ie^{-\nu  Z}A_p^{(r)},\hskip .5cm A_{s,i}=e^{-\nu  Z}A_{s,i}^{(r)}.
\end{eqnarray}
In the new variables Eqs. (\ref{umpa}),  (\ref{umpb}) take the form (compare with Ref. \cite{cullen})
 \begin{subequations}
\begin{alignat}{4}
\frac{dA_p^{(r)}}{d\tilde{Z}}&=\mu_p A_s^{(r)}A_i^{(r)}  \label{ra}\\
\frac{dA_{s,i}^{(r)}}{d\tilde{Z}}&=-\mu_{s,i}A_p^{(r)}A_{i,s}^{(r)},\label{rb}\end{alignat}
\end{subequations}
where
\begin{eqnarray}
d\tilde{Z}\equiv  e^{-\nu Z} dZ,
\end{eqnarray}
that is
\begin{eqnarray}
\tilde{Z}=K-\frac{1}{\nu}e^{-\nu Z},
\end{eqnarray}
where $K$ is an arbitrary constant.

\begin{figure}
\includegraphics[width=.8\columnwidth]{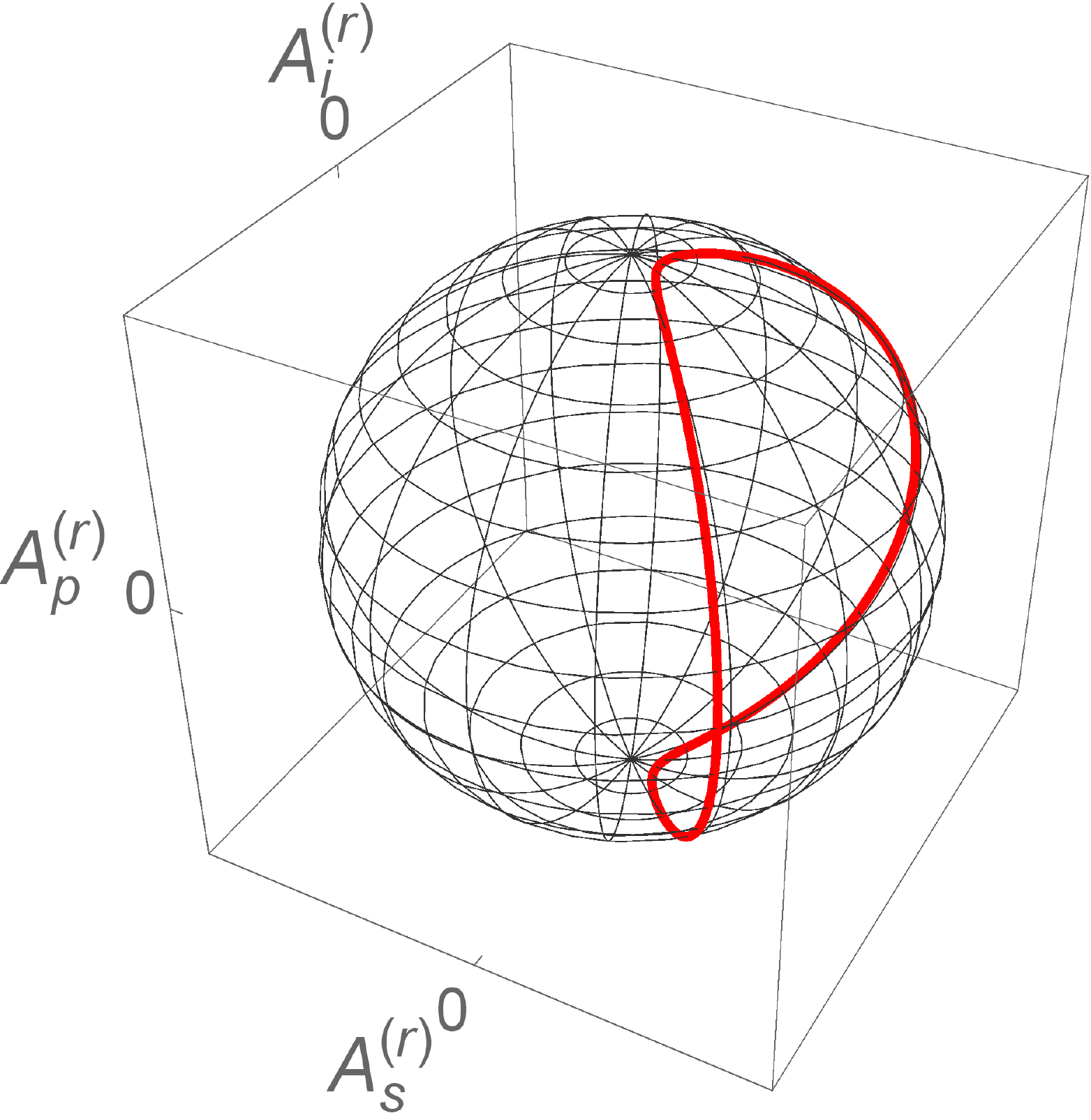}
\caption{The solution (\ref{good2}) plotted in the variables  $A_{\alpha}^{(r)}$.  We have chosen   $\mu_i/\mu_s=1.2$ and $\theta=.1$  for  $\phi=0$. }
 \label{cul2}
\end{figure}

We'll consider only the real solutions, postponing the general analysis until later.
From (\ref{ra}), (\ref{rb}) follows
\begin{eqnarray}
\label{consta}
\frac{1}{2}\frac{d}{d\tilde{Z}}\left[{A^{(r)}_p}^2+{A^{(r)}_s}^2+{A^{(r)}_i}^2\right]
\nonumber\\
=\left(\mu_p  -\mu_s-\mu_i\right)A_p^{(r)}A_s^{(r)}A_i^{(r)}.
\end{eqnarray}
Due to (\ref{mu}) the r.h.s. of (\ref{consta}) is equal to zero, hence
\begin{eqnarray}
\label{const}
{A^{(r)}_p}^2+{A^{(r)}_s}^2+{A^{(r)}_i}^2=\text{const},
\end{eqnarray}
and we can look for a solution in the form (remember the spherical coordinates)
\begin{subequations}
\begin{alignat}{4}
A_p^{(r)}&=B\cos\theta  \label{mpa}\\
A_s^{(r)}&=B\sin\theta\cos\phi \label{mpb}\\
A_i^{(r)}&=B\sin\theta\sin\phi,      \label{mpc}
\end{alignat}
\end{subequations}
where $B$ is an arbitrary amplitude. For the new variables $\theta$ and $\phi$ we obtain equations
\begin{subequations}
\begin{alignat}{4}
\frac{d\theta}{d\tilde{Z}}&=-\frac{B}{2}\mu_p \sin\theta \sin 2\phi \label{pa}\\
\frac{d\phi}{d\tilde{Z}}&=-\frac{B}{2} \cos\theta\left(\mu_p\cos 2\phi -\mu_s+\mu_i\right).\label{pb}
\end{alignat}
\end{subequations}
Dividing one of the equations to the other we obtain
\begin{eqnarray}
\label{good}
\frac{d\theta}{\tan\theta }=\frac{\mu_p\sin 2\phi d\phi}
{\mu_p\cos 2\phi-\mu_s+\mu_i}.
\end{eqnarray}
Equation (\ref{good}) can be easily integrated and we obtain
\begin{eqnarray}
\label{good2}
\sin^2\theta =\frac{D}{\mu_p\cos 2\phi-\mu_s+\mu_i},
\end{eqnarray}
where $D$ is the integration constant. The solution (\ref{good2}), plotted in the variables  $A_{\alpha}^{(r)}$,  is graphically presented in Fig. \ref{cul2}.
In the variables   $A_{\alpha}$  the
sphere contracts uniformly.

Equation (\ref{good2}) is even simpler than it looks.Taking into account (\ref{const}), it can be written down as
\begin{eqnarray}
(\mu_s-\mu_i){A_p^{(r)}}^2+\mu_p\left({A_s^{(r)}}^2-{A_i^{(r)}}^2\right)=\text{const}.
\end{eqnarray}
The latter equation follows immediately from (\ref{ra}), (\ref{rb}).

There also exists a singular solution of (\ref{pa}),  (\ref{pb}), where the constant $\phi$ is given by the equation
\begin{eqnarray}
\label{good3}
\mu_p\cos 2\phi-\mu_s+\mu_i=0,
\end{eqnarray}
and $\theta(\tilde{Z})$, found from  (\ref{pa}), is given by the equation
\begin{eqnarray}
\label{tt}
\cos \theta=\tanh\left(\frac{B}{2}\mu_p \sin 2\phi\cdot \tilde{Z}\right ).
\end{eqnarray}
Equation (\ref{tt}) shows that when $ \tilde{Z}\to\infty$, either $\theta\to 0$ or $\theta\to\pi$, depending upon the sign of $\sin 2\phi$.

\begin{figure}
\includegraphics[width=.8\columnwidth]{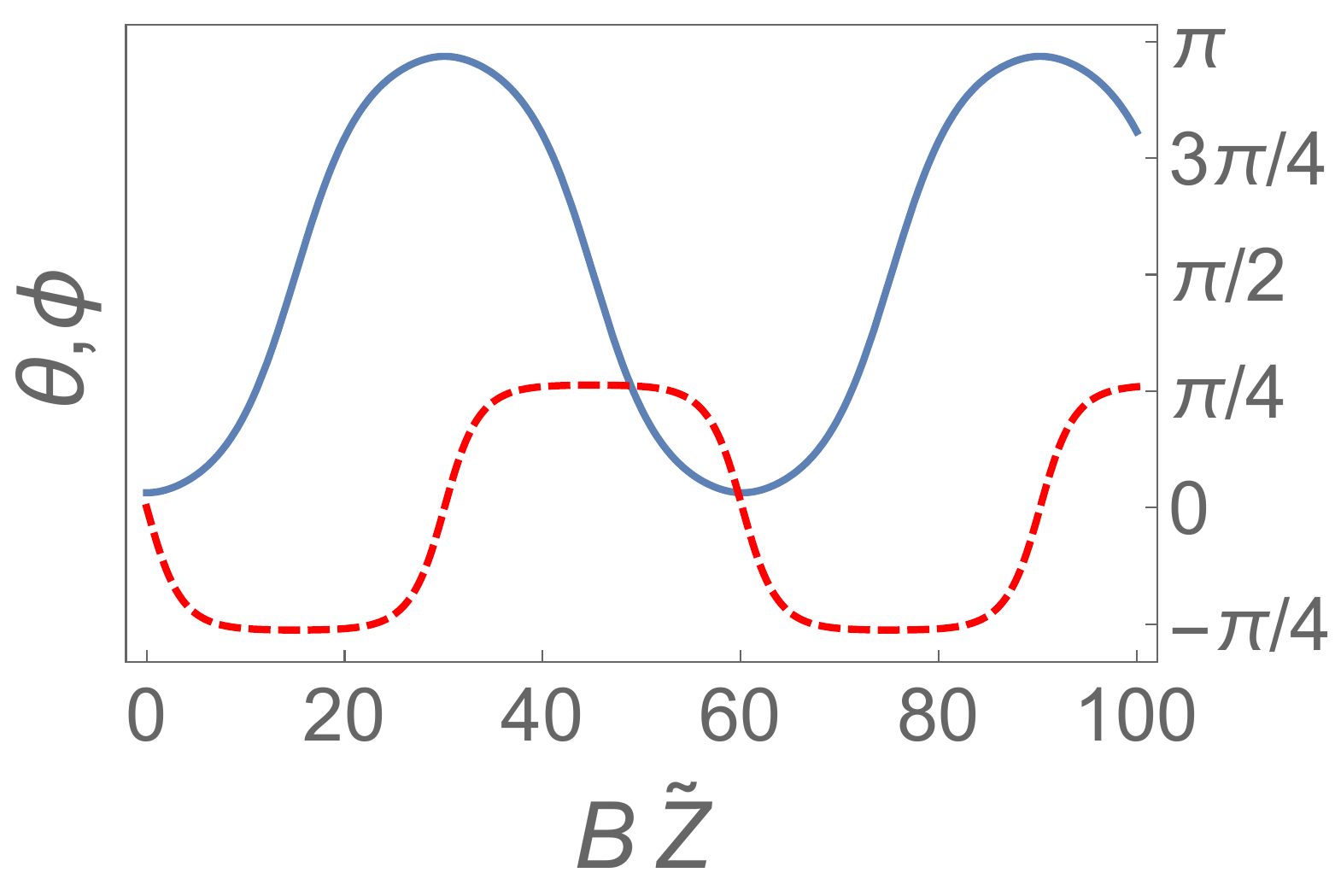}
\caption{Numerical solution of Eqs. (\ref{pa}), (\ref{pb}) for $\theta$ (blue solid line) and $\varphi$ (red dashed line). We have chosen   $\mu_s=.2$, $\mu_i=.24$, $\theta(0)=.1$ and  $\varphi(0)=0$ (no idle signal at $\tilde{Z}=0$). }
 \label{cul}
\end{figure}

However, we are mostly interested not in the dependence of
one angle variable upon the other, but in their dependence
upon $\tilde{Z}$. This dependence can be found by  numerical integration of Eqs. (\ref{pa}), (\ref{pb}) and  is  illustrated  in Fig. \ref{cul}. The left side of the Figure, with $\theta$ increasing from nearly 0, to $\pi/2$, demonstrates the parametric amplification (provided the losses are not too big).

\section{4 waves mixing}
\label{cubic}

\subsection{Coupled equations for the amplitudes}

Now let us consider the case $\varphi_0=0$. In this case, expanding the sine function in the r.h.s. of (\ref{co}) in Taylor series with respect to phases difference and keeping the first two nonzero terms of the expansion we have
\begin{eqnarray}
n.l.\left[\sin\left(\varphi_{n}\right)\right]
=-\frac{1}{6}\varphi_{n}^3.
\end{eqnarray}
We again consider the superposition of three waves, but
the frequencies this time  satisfy the relation
\begin{eqnarray}
\omega_s+\omega_i= 2\omega_p.
\end{eqnarray}
 Keeping only the terms which have
$\omega_{p,s,i}t$ time dependence  we obtain
\begin{eqnarray}
\varphi_{n}^3=
\frac{3}{8}\sum_{\alpha,\beta}\left(2-\delta_{\alpha\beta}\right)
A_{\alpha}\left|A_{\beta}\right|^2e^{i[k_{\alpha}n-\omega_{\alpha}t]}\nonumber\\
+\frac{3}{4} A_p^*A_sA_ie^{i[(k_s+k_i-k_p)n-\omega_pt]}
+\frac{3}{8} A_p^2A_i^*e^{i[(2k_p-k_i)n-\omega_st]}\nonumber\\
+\frac{3}{8}A_p^2A_s^*
e^{i[(2k_p-k_s)n-\omega_it]}+c.c..
\end{eqnarray}

Finally, collective all the perturbative terms in (\ref{co}) we
obtain coupled equations for the  amplitudes
\begin{subequations}
\label{coco1}
\begin{alignat}{4}
\frac{dA_p}{dZ}+\nu_pA_p&=i\gamma_p\left(|A_p|^2+2|A_s|^2
+2|A_i|^2\right)A_p\nonumber\\
&+2i\eta_pA_p^*A_sA_ie^{i\Delta k\cdot Z}   \label{coco1a} \\
\frac{dA_{s,i}}{dZ}+\nu_{s,i}A_{s,i}&=i\gamma_{s,i}\left(2|A_p|^2
+|A_{s,i}|^2+2|A_{i,s}|^2\right)A_{s,i}\nonumber\\
&+i\eta_{s,i}A_p^2A_{i,s}^*e^{-i\Delta k\cdot Z}, \label{coco1b}
\end{alignat}
\end{subequations}
where $\nu_{\alpha}$ is given by (\ref{nu}) with $\cos\varphi_0$ being put equal to 1, and
\begin{subequations}
\label{ga2}
\begin{alignat}{4}
\gamma_{\alpha}&=\frac{ \tan (k_{\alpha}/2)}{2\omega_{\alpha}^2} \sin^2\left(\frac{k_{\alpha}}{2}\right)\label{ga}\\
\eta_p&=\frac{ \tan (k_p/2)}{2\omega_p^2} \sin^2\left(\frac{k_s+k_i-k_p}{2}\right)\\
\eta_{s,i}&=\frac{ \tan (k_{s,i}/2)}{2\omega_p^2}
\sin^2\left(\frac{2k_p-k_{i,s}}{2}\right) \\
\Delta k&=k_s+k_i-2k_p
\end{alignat}
\end{subequations}

Comparing Eqs. (\ref{coco1})  with the appropriate equations from our previous publication \cite{kogan3} we can only repeat what was said in the very end of Section \ref{adm}.

\subsection{Weak signal}

In  the small (relative to the pump) signal and idler approximation   ($|A_s|,|A_i|\ll |A_p|$),
 the
equation for $A_p$ becomes decoupled from the other two and takes the form
\begin{eqnarray}
\label{ne}
\frac{dA_p}{dZ}+\nu_pA_p=\frac{i}{2}\gamma_p|A_p|^2A_p.
\end{eqnarray}

It is interesting to compare the result following from equation (\ref{ne}) with
the nonlinear dispersion law obtained in our previous publication \cite{kogan}.
There we considered equation
\begin{eqnarray}
\label{cop}
\frac{d^2\varphi_n}{d\tau^2}=
\sin \varphi_{n-1}-2\sin \varphi_n
+\sin \varphi_{n+1}
\end{eqnarray}
and obtained in the up to the second order with respect to the amplitude approximation the solution
\begin{eqnarray}
\varphi_n(\tau)=a\cos[kn-\omega(k)\tau]
\end{eqnarray}
with the dispersion law
\begin{eqnarray}
\label{old}
\omega(k)=2\left(1-\frac{a^2}{16}\right)\sin\left(\frac{k}{2}\right).
\end{eqnarray}

In the absence of dissipation, the solution of (\ref{ne}) can be taken as
\begin{eqnarray}
A_p(Z)=ae^{i\gamma_pa^2/2Z},
\end{eqnarray}
where $a$ is a real constant. Substituting this solution into (\ref{combin}) and taking into account (\ref{ga}) we obtain
\begin{eqnarray}
\label{add}
\varphi_n(\tau)=a\cos[(k+a^2\tan(k/2)/8)n-\omega\tau],
\end{eqnarray}
where $\omega=2\sin(k/2)$. let us introduce
\begin{eqnarray}
k'=k+a^2\tan(k/2)/8.
\end{eqnarray}
Then up to the lowest order with respect to $a^2$ corrections
\begin{eqnarray}
k=k'-a^2\tan(k'/2)/8,
\end{eqnarray}
and the dispersion law can be presented as
\begin{eqnarray}
\label{disper}
\omega=2\sin[k'/2-a^2\tan(k'/2)/16].
\end{eqnarray}
Expanding (\ref{disper}) in series with respect to $a^2$ and keeping the terms up to the linear order with respect to  $a^2$  we obtain
\begin{eqnarray}
\label{new}
\omega(k')=2\left(1-\frac{a^2}{16}\right)\sin\left(\frac{k'}{2}\right).
\end{eqnarray}
which coincides with Eq. (\ref{old}).

Note that in our previous publications \cite{kogan} the pump wave hire harmonics,
which exist simultaneously with the main one, due to nonlinearity of the system,
are also considered. However there influence on the signal amplitude would be an effect of the higher order with respect to the parameter $A_p^2$, than
the influence of the main harmonic.

\subsection{Threshold for the parametric amplification}

Equations (\ref{coco1b}) in the weak signal approximation can be written down as
\begin{eqnarray}
\label{ho9}
\frac{dA_{s,i}}{dZ}+\left(\nu_{s,i}-i\gamma_{s,i}|A_p|^2\right)A_{s,i}
=i\eta_{s,i}A_p^2A_{i,s}^*e^{-i\Delta k\cdot Z}.
\end{eqnarray}
In the framework of the local approximation,  we treat $A_p$ in the r.h.s. of Eqs. (\ref{ho9}) as a constant, so we recover
 (\ref{soli}), where $K$   this time is one of the roots of the characteristic polynomial
\begin{eqnarray}
\label{sol2b}
P(K)\equiv K^2+(\nu_s+\nu_i)K
+(\nu_s-2i\gamma_s|A_p|^2-i\Delta k/2)\nonumber\\
\cdot(\nu_i+2i\gamma_i|A_p|^2+i\Delta k/2)-\eta_s\eta_i\left|A_p\right|^4.
\end{eqnarray}
We again find the boundary between the parametric amplification and no amplification  by demanding that $K$ for that boundary is purely imaginary. So,
after simple algebra, we obtain
the condition for the parametric amplification:
\begin{eqnarray}
\label{4}
\eta_s\eta_i\left|A_p\right|^4>\nu_s\nu_i\left\{1+\frac{\left[\Delta k+2(\gamma_s+\gamma_i)\left|A_p\right|^2\right]^2}{(\nu_s+\nu_i)^2}\right\}.
\end{eqnarray}
Here we can only repeat what was said after Eq. (\ref{3}).
Again, we bring back the coordinate dependence of $A_p$, given by Eq. (\ref{ne})
and  consider (\ref{4}) as a local condition for the parametric amplification.

\section{Discussion}

The present paper generalized the previously obtained results
on parametric amplification in JTL \cite{yaakobi,zorin,kogan} to the discrete model of the line.  The discrete character of the line leads to the renormalization of the parameters of the coupled amplitude equations.

We formulate the local approximation which allows explicitly find the
threshold for parametric amplification in the case of weak signal.
In addition we present the exact particular solution for the coupled equations for the pump, signal and idler amplitudes in the particular case of small wave vectors and
present the solution graphically.
Finally, we would like to mention that
in real case of
the line of finite length, the frequency dependence of the line impedance leads to matching
problems and hence to the wave reflection (e.g., \cite{kern}). However we leave the study of the problem until later.

\section*{Acknowledgment}
I am grateful to  T. H. A. van der Reep for the discussion.
I am also very grateful to Donostia International Physics Center (DIPC)
for the hospitality during my visit, when the paper was finalised.

\end{document}